\documentclass[useAMS,usenatbib,twocolumn]{mn2e}
\usepackage[]{graphicx}
\usepackage{latexsym}
\usepackage{amssymb}
\usepackage{float}
\usepackage{ifthen}
\usepackage{url}

\newcommand {\hI} {{H}$_\mathrm{I}$}
\newcommand {\hII} {{H}$_\mathrm{II}$}
\newcommand {\ha} {H$\alpha$}

\newcommand {\kms} {\,km\,s$^{-1}$\,}

\newcommand{\Min}{${}^{\prime}$}
\newcommand{\Sec}{${}^{\prime\prime}$}

\newcommand{\FM} {\texttt{\textsc{FaNTOmM}}}

\title[Improved 3D FP Data Reduction Techniques]{Improved 3D Fabry-Perot Data Reduction Techniques\thanks{Based on observations collected at the European Southern Observatory, La Silla, Chile and at the Canada-France-Hawaii Telescope (CFHT) which is operated by the National Research Council of Canada, the Institut des Sciences de l'Univers of the Centre National de la Recherche Scientifique of France, and the University of Hawaii.}}
\author[O. Daigle et al.]{O. Daigle$^{1}$\thanks{E-mail:
odaigle@astro.umontreal.ca}, C. Carignan$^{1}$, 
O. Hernandez$^{1,2}$, L. Chemin$^{1,3}$ and P. Amram$^{2}$\\
$^{1}$Observatoire du mont M\'egantic, LAE, Universit\'e de Montr\'eal, C. P. 6128 succ. centre ville, Montr\'eal, Qu\'ebec, Canada H3C 3J7\\
$^{2}$Observatoire Astronomique de Marseille Provence, LAM, 2 place Le Verrier, F-13248 Marseille Cedex 04, France\\
$^{3}$Observatoire de Paris, section Meudon, GEPI, CNRS UMR 8111 et Universit\'e 
Paris 7, 5 place J. Janssen, 92195 Meudon Cedex,\\France.\\}
\begin{document}

\date{Accepted . Received ; in original form }

\pagerange{\pageref{firstpage}--\pageref{lastpage}} \pubyear{2006}

\maketitle

\label{firstpage}

\begin{abstract}
Improved data reduction techniques for 3D data cubes obtained from Fabry-Perot integral field spectroscopy are presented. They provide accurate sky emission subtraction and adaptive spatial binning and smoothing. They help avoiding the effect analogous to the beam smearing, seen in \hI\, radio data, when strong smoothing is applied to 3D data in order to get the most extended signal coverage. The data reduction techniques presented in this paper allow one to get the best of both worlds: high spatial resolution in high signal-to-noise regions and large spatial coverage in low signal-to-noise regions.

\end{abstract}

\begin{keywords}
methods: data analysis $-$ techniques: 3D spectroscopy.
\end{keywords}

\section{Introduction}

In recent years, many \ha\, kinematical surveys have been done in photon counting mode (\citeauthor{2003SPIE.4841.1472H} \citeyear{2003SPIE.4841.1472H}, \citeauthor{2002PASP..114.1043G} \citeyear{2002PASP..114.1043G}) using a Fabry--Perot (FP) interferometer on large galaxies' samples (e.g. \citeauthor{2002Ap&SS.281..393A} \citeyear{2002Ap&SS.281..393A}, \citeauthor{2003AJ....126.2635M} \citeyear{2003AJ....126.2635M}, \citeauthor{2003MNRAS.343..819R} \citeyear{2003MNRAS.343..819R}, \citeauthor{2005A&A...436..469C} \citeyear{2005A&A...436..469C}, \citeauthor{2005MNRAS.360.1201H} \citeyear{2005MNRAS.360.1201H}, \citeauthor{2005sdlb.procP..47M} \citeyear{2005sdlb.procP..47M}, \citeauthor{2006MNRAS.367..469D} \citeyear{2006MNRAS.367..469D}). For galaxies, these observations may present differences nearly as large as two orders of magnitude in signal-to-noise ratio (SNR) between starburst regions and diffuse (inter-arm) regions. Even after hours of integration, which means, in fact, about 5 minutes per FP channel, the signal in diffuse areas is often too weak to give reliable results.

Since the SNR needed depends on the moments of the line profile that has to be analysed (flux, velocity, dispersion, skewness, kurtosis, ...), a typical solution to this problem is to apply a spatial weighted averaging (or smoothing) to correlate neighbouring data points. Doing so, the spatial area over which the line profiles can be analysed can be greatly increased at the price of a degraded spatial resolution, hiding sometimes structures that require high resolution to be rendered. Thus, to make it possible to resolve small kinematical details in the galaxy structure and to have the most extended coverage possible, it is necessary to create different velocity maps with different levels of smoothing. A typical approach is to create a non smoothed map, a map smoothed with a 3x3 Gaussian kernel and another one convolved with a 6x6 Gaussian kernel. Then, the data analysis must be done three times. Moreover, lowering the spatial resolution involves a flux contamination by extending erroneously signal from high SNR pixel areas to low SNR pixel zones.

\begin{figure}
\includegraphics[width=0.48\textwidth]{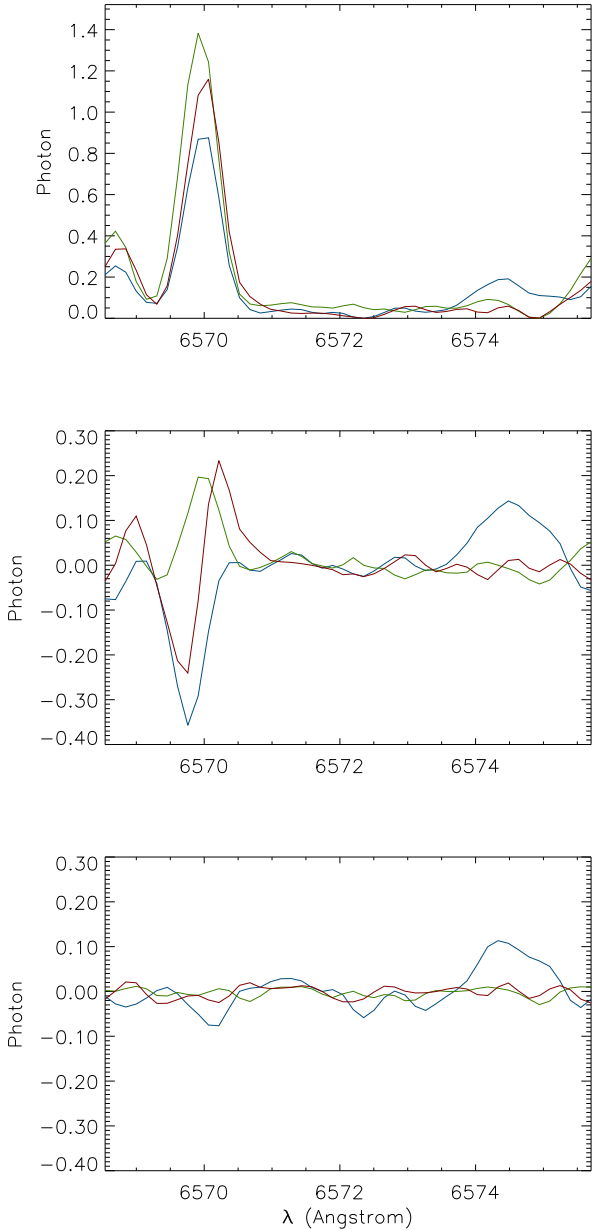}
\caption{Integrated spectra (continuum subtracted) of three different sky regions as seen in the data cube of the galaxy \mbox{NGC 5194}. Each region is roughly composed of 500 pixels. The plots are normalised accordingly. The pixel size on the sky is 2.56 $\mathrm{arcsec^2}$. The first order of the spectra is between 6569.0 and 6576.3 \AA. There are 48 spectral bins for every spectrum. \mbox{\textbf{Top} :} Before sky subtraction. The spectra discrepancies are due to inhomogeneous interference filter transmission. \mbox{\textbf{Middle} :} After subtracting a median spectrum from the data cube. \mbox{\textbf{Bottom} :} After subtracting a sky cube from the data cube. The feature around 6574.5\AA\, comes from the sky emission line at 6596.64\AA\, that is not perfectly suppressed. }
\label{figure_flat}
\end{figure}

Given these characteristics, \ha\, FP observations of galaxies would greatly benefit from an adaptive smoothing scheme that would be a function of the SNR of the resolution element. Similar work has been done on integral-field spectroscopy (IFS) (\citeauthor{2003MNRAS.342..345C} \citeyear{2003MNRAS.342..345C}) with Voronoi tessellation. The main aim of their work was to bin 3D data so that the resulting SNR of each bin would tend towards a single value. The algorithm they developed has been worked on and adapted to scanning FP data, as described in section \ref{section_binning}. Results are presented in section \ref{section_results}.

However, while adaptive smoothing can allow increasing the SNR in weak emission regions without decreasing the spatial resolution in high SNR areas, in FP mode other aspects of the data reduction must be considered prior to smoothing. Section \ref{section_sky_emission} will show that the sky emission spectrum can not be considered homogeneous throughout the data cube and an efficient way to subtract it will be provided.



\section{Sky emission in FP data cubes}
\label{section_sky_emission}
\subsection{Sky emission inhomogeneities}
\label{sky_inhomogeneities}

\begin{figure}
\includegraphics[width=0.48\textwidth]{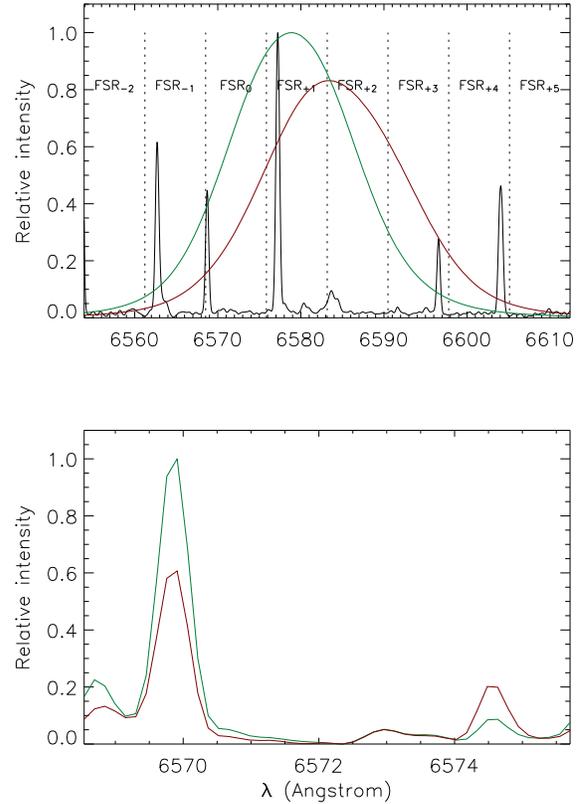}
\caption{\textbf{Top} : Transmission curves at two different positions on the interference filter used to observe \mbox{NGC 5194} (red and green lines). The Ultraviolet and Visual Echelle Spectrograph (UVES; \citeauthor{2000SPIE.4008..534D} \citeyear{2000SPIE.4008..534D}) sky spectrum around the central wavelength of the filter is shown as the black line. The spectrum resolution have been degraded to the one obtained in the data cube of \mbox{NGC 5194} (R$\sim$18600). The FP etalon Free Spectral Ranges for this galaxy are also shown. \textbf{Bottom} : The sky spectrum as seen in a data cube considering the different transmission curves of the filter. The sky spectrum present in every FSR of the FP etalon have been folded back into $\mathrm{FSR_0}$.}
\label{figure_filtre}
\end{figure}

Night sky emission lines are very prominent in the red part of the visible spectrum. Around \ha\, at rest, geocoronal $\mathrm{H_{\alpha 0}}$\, (6562.8 \AA) and strong OH sky emission lines (e.g.: 6553.6 \AA, 6568.8 \AA, 6577.2 \AA, 6577.4 \AA) contaminate the spectrum. When \ha\, is redshifted, weak to strong OH lines (e.g.: 6583.41 \AA, 6596.64 \AA, 6603.99 \AA, 6604.28 \AA) continue to contaminate the spectrum at different levels (for a detailed sky emission lines atlas, see \citeauthor{1996PASP..108..277O} \citeyear{1996PASP..108..277O}). Although the wavelength and relative intensities of these lines are well documented, it is noticeable that they are not homogenous across all the field of view (see top panel of Figure \ref{figure_flat}).

The strength of the OH emission varies throughout an observation :
\begin{itemize}
\item The OH radicals' column density changes with the airmass;
\item The OH emission changes as the line of sight crosses different OH radicals layers;
\item Radiation (sunrise, sunset, moon) excites the OH radicals.
\end{itemize} 

These variations affect large scale OH emission structures. Over time, since the FP etalon is scanned at least ten times during an observation (and sometimes up to 50 times), these variations are averaged out. Thus, the main contribution to the inhomogeneities of the sky spectrum is not the OH emission variation.

Interference filters are placed in front of the FP etalon to restrict the wavelength interval seen by the etalon. These filters have a typical profile width of 15\AA. Their central wavelength is chosen to match the systemic velocity of the galaxy. By scanning the filters with a spectrophotometer, it has been realised that the filter's transmission shape changes spatially. Since these filters are placed in a converging beam (rather than into a pupil), their transmission inhomogenities are visible in the data cube (Figure \ref{figure_filtre}). The dissimilarity of the sky spectrum observed in the data cubes and the ones modelled agree (bottom panel of Figure \ref{figure_filtre} and top panel of Figure \ref{figure_flat}). Thus, one cannot assume that the sky emission spectrum seen in a single FSR of the FP etalon will be homogeneous across the field of view as long as the interference filters transmission is not perfectly uniform.

The subtraction of the sky emission is an important process. As it will be described in section \ref{section_binning}, the kinematical information is extracted by looking at only one emission line. An improperly removed sky emission will cause inaccurate velocities to be found as the amplitude of the sky residuals may be greater than the galaxy's signal. In order to properly subtract it, one must use a process finer than just subtracting a median spectrum (middle panel of Figure \ref{figure_flat}).

\begin{figure}
\includegraphics[width=0.48\textwidth]{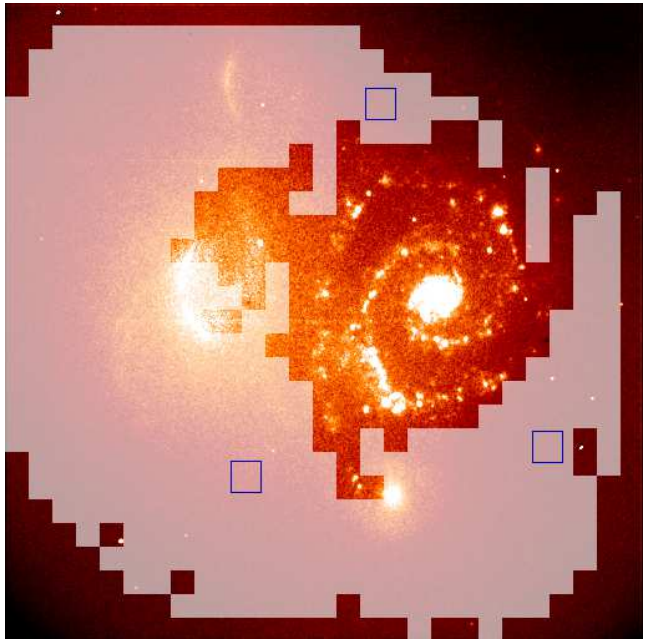}
\caption{Spectrally summed data cube of \mbox{NGC 5194} observed in \ha\, light. The lighter sections of the image are those considered as sky-dominated by the algorithm described in Section \ref{sky_cube}. The boxed blue regions are those whose spectra are shown in Figure \ref{figure_flat}. Note that the galaxy's companion, \mbox{NGC 5195}, lacks \ha\, emission and has not been considered as galaxy-dominated. The bright halo on the left side of \mbox{NGC 5194} is a ghost coming from the bright core of the galaxy. Since this ghost contains emission lines from the galaxy, some parts of it are considered as galaxy-dominated. Ghost subtraction is a tedious task and it is being worked on.}
\label{figure_ngc5194_sky}
\end{figure}

\subsection{Sky cube subtraction}
\label{sky_cube}

To account for the differences in sky emission seen in the data cube, one must construct a sky cube and subtract it from the data cube in order to properly remove the sky emission. To do so, the data cube must be divided into two parts: the galaxy-dominated regions and the sky-dominated regions. Then, a sky cube may be constructed by fitting the sky-dominated regions and interpolating the sky spectrum in the galaxy-dominated regions. The most challenging part of the problem is to disentangle the sky-dominated regions from the galaxy-dominated ones.

The galaxy's emission that is redshifted by rotation moves spectrally across the data cube. Thus, the spectrum that is the most constant across the data cube is the sky spectrum. By taking the spectrum of the data cube (the median of the spectrum that is above the continuum), one gets an \textit{estimate} of the sky emission spectrum structure. Then, the data cube is spatially binned to get a fairly good SNR on the sky emission. An optimised target SNR was determined to be $\sim40$. Depending on the strength of the night sky emission, the total integration time and the pixel size on the sky, the size of the bins ranges from $\sim250$ to $\sim625$ square pixels, giving an angular size on the sky ranging from $\sim100$ to $\sim1000$ $\mathrm{arcsec^2}$. Then, the spectrum of these bins is cross-correlated with the median spectrum of the data cube. However, as discussed in section \ref{sky_inhomogeneities}, the sky spectrum seen in the data cube changes from one region to another. Thus, in order to determine the sky-dominated regions, one must chose a correlation factor that is lower than 1.0. Since the apparent wavelength shift of the spectra is typically smaller than a channel ($\sim$0.2\AA), a correlation factor of 0.9 gives a good approximation of the sky-dominated regions. To better disentangle the galaxy-dominated regions from the sky-dominated ones (in cases where the galaxy emission falls directly on a sky emission line), the emission amplitude of sky-dominated regions must be within $\pm50$ percent of the median spectrum. Once the sky-dominated regions are determined (see Figure \ref{figure_ngc5194_sky}), a $\mathrm{4^{th}}$ degree irregular surface fit is made to interpolate the sky spectrum into the galaxy-dominated regions. Every sky-dominated region is given the same weight.

Finally, this sky cube is subtracted from the original data cube. The result of this process is shown in the bottom panel of Figure \ref{figure_flat}.


\section{Adaptive binning of 3D data cubes}
\label{section_binning}
When reducing a data cube, the main objective should be to get the best possible spatial and spectral resolutions. The spatial resolution is generally limited by the seeing if the physical sampling of the seeing is sufficient. However, it is not as straightforward when considering the spectral resolution, which depends on the SNR. While it is an easy task to reach the theoretical spectral resolution in bright regions, it becomes less obvious in low SNR areas.

Doing a spatial averaging by correlating neighbouring pixels is a well known method to enhance the signal coverage. However, in many cases, this solution is unsatisfactory since it degrades the spatial resolution even in the brightest emission line regions where it is not necessary, hiding structures that require high spatial or spectral resolution to be rendered. Moreover, this smoothing induces not only a spatial pollution but also a spectral one as it transports the flux from the most intense regions towards the weaker emission areas, completely drowning the signal in these areas. For instance, the beam smearing effect in \hI\, data (\citeauthor{2000ApJ...531L.107S} \citeyear{2000ApJ...531L.107S}) produces an underestimation of the \hI\, rotation velocity at small galactic radii where the velocity gradient is the greatest.

\ha\, emission intensities in galaxies display a very large dynamic range (several orders of magnitude), from the brightest compact and discrete \hII\, regions to the weaker diffuse \ha\, emission in the inter-arm regions. The higher spatial and spectral resolutions can usually be reached from the \hII\, regions without any smoothing. On the other hand, the contrast may be very sharp between a bright \hII\, region and a weak diffuse emission area.

It is thus relevant to introduce the concept of spatial smoothing adapted to a given SNR on the same data cube allowing maximum spatial resolution in regions of strong emission (e.g. \hII\, regions) and maximum signal coverage in regions of weak emission (e.g. inter-arm regions). Such an approach was recently used by Cappellari and Copin (\citeyear{2003MNRAS.342..345C}), working on IFS 2D data. The aim of their work was to bin radial velocity maps according the spectra SNR. According to them, the ideal binning scheme should give bins that satisfy the following requirements :

\begin{itemize}
\item Topological requirement : The bins should properly cover the sky region, without holes or overlaps.
\item Morphological requirement : The bins should be as round as possible to optimise the spatial resolution.
\item Uniformity requirement : The resulting SNR of every bin should tend towards a single value. A minimum value should be required and no more attempts should be done to accrete pixels on a bin that already have the chosen SNR.
\end{itemize}

The uniformity requirement is easily fulfilled when the resulting SNR of a bin is computed by mean of
$$SNR_{binned} = \frac{\sum\limits_{i}Signal_{i}}{\sqrt{\sum\limits_{i}Noise_{i}^{2}}},$$
where $Signal_{i}$ and $Noise_{i}$ are the signal and noise value of the single $i$th pixel composing a bin. However, as the different spectral components of every pixel are not taken into account, it is impossible to predict the real SNR of a spectrum after its addition to another one. This binning scheme also assumes that reaching a greater SNR is simply a matter of adding more pixels to the bin. Nevertheless, as spatial resolution is limited, accreting pixels that are too distant leads to adding pixels that are not correlated together as they come from different Line Of Sight Velocity (LOSV) regions. The binning would be more accurate if the SNR of a bin would be recalculated each time a new pixel is added to the bin. As the addition of pixels that come from different LOSV regions will not increase the SNR of an emission line, this will avoid merging structures that are not related.



\begin{figure*}
\includegraphics[width=0.99\textwidth]{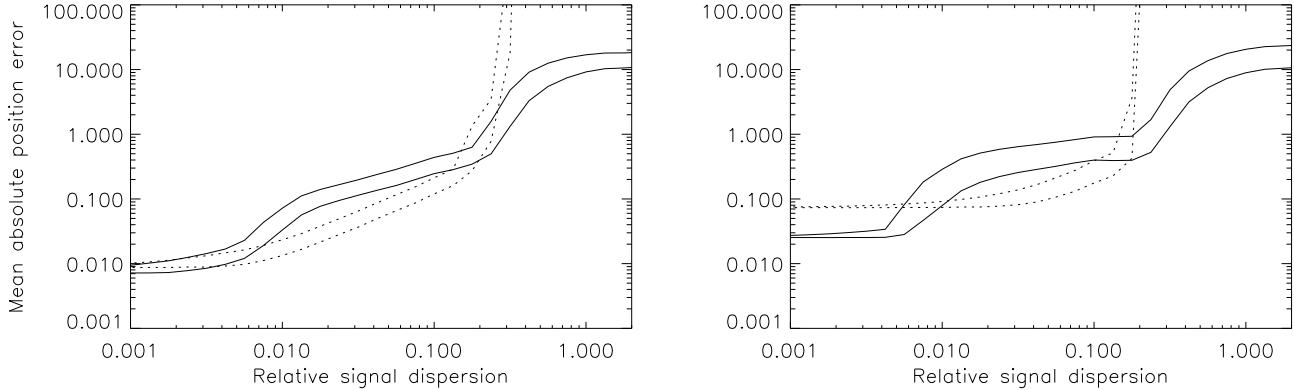}
\caption{Comparison of the emission line position determination algorithm (plain curve) and a Gaussian fitting algorithm (dotted curve). The signal dispersion relative to the emission line height is plotted on the x axis. The lower line is the mean absolute position error (in FP channel) and the upper line is the mean absolute position error plus the position error dispersion ($1 \sigma$). \mbox{\textbf{Left} :} Results without sky residuals. \mbox{\textbf{Right} :} Results with sky residual. Sky residuals were $0.2$ the amplitude of the emission line.}
\label{figure_snr}
\end{figure*}

\subsection{Finding the emission lines}
\label{section_finding_emission}
The determination of the SNR of the emission lines is the key of the binning algorithm. Before any attempt is made at determining the SNR of an emission line, we first have to find where the emission line is located.

The boundaries of the emission lines are determined by a two-pass algorithm. First, the mean signal level of the spectrum is found. Then, the highest channel of the spectrum is taken as the starting point of the emission line. If the value of the two channels on both sides of the maximum are greater than 75 per cent of the mean spectrum value, this point is considered to belong to the emission line (the threshold value of 75 per cent is chosen to match the amplitude of the two channels each side of the maximum of an Airy function that is properly (Nyquist) sampled). If not, the algorithm will skip to the next maximum of the spectrum until one is found that fulfil the restriction. This maximum is the starting point. This mechanism has been put in place when processing CCD data. A cosmic ray could have been the maximum point and it must be avoided to look for an emission line around it. In photon counting, where the observation is multiplexed and the camera is operated at a high frame rate, this can not happen. This mechanism also helps finding emission lines in really low SNR spectra.

Then, by walking both sides of the maximum, the first points that pass below the median flux and have a change in the sign of the slope are found. This delimitates the rough emission line. If a change in the sign of the slope is never encountered, the first point that falls below the median is taken as the boundary. Using those boundaries, the continuum is computed by taking the mean value of the channels that fall outside the emission line. The dispersion of that first continuum is computed as follows
$$\sigma = \sqrt{\frac{\sum\limits_{i}^{i=n-1}y^{2}_{i} - \left(\sum\limits_{i}^{i=n-1}y_{i}\right)^{2}n}{n-1}}.$$



The second pass involves starting back from the maximum channel previously found and walking both sides of it to find the first points that pass below the continuum plus dispersion, but if a change in the sign of the slope is encountered in the $2 \sigma$ range of the continuum, this point is taken as the boundary. This allows a better avoidance of the sky residuals that could still pollute the galaxy's spectra. The boundaries found are taken as the true boundaries of the emission line. 

Now that the emission line is isolated, it is possible to extract the parameters of the spectrum. The continuum is computed by taking the mean signal outside the emission line and its dispersion is calculated. The centre of the emission line is simply computed as the intensity weighted centroid of the spectral bins falling into the emission line boundaries, with the continuum subtracted. The amount of spectral bins used depends on the width of the emission line.

Figure \ref{figure_snr} shows how this \textit{selective} intensity weighted mean algorithm performs as compared to a Gaussian fitting algorithm (\textsc{gaussfit} in \textsc{IDL}). The simulated spectra were made of a Gaussian convolved with an Airy function. Noise with a normal distribution was then added to the spectra to account for the many kinds of noise that can be encountered. For spectra where the signal dispersion relative to the amplitude of the emission line was $> 0.2$, the Gaussian fitting algorithm diverges completely. For less noisy spectra ($0.004 < \sigma_{relative} < 0.2$), the Gaussian fitting algorithm performs slightly better. However, the additional error induced by the selective intensity weighted mean is less than 0.15 channel.

In the right panel of Figure \ref{figure_snr}, the same comparison is made but it uses a spectrum that is polluted with a sky residual that has an amplitude that is 20 per cent that of the emission line. As expected, the Gaussian fitting algorithm performs better for low noise spectra ($0.01 < \sigma_{relative} < 0.1$), with an absolute error that is 0.2 channel lower. However, the Gaussian fitting algorithm is still confused by noisy spectra and diverges.

It was decided to use the selective intensity weighted mean algorithm since it performs very well for noisy spectra. This signal regime is usually seen in very diffuse areas and the ability to resolve kinematics in these regions dictates how extended is the signal coverage. 


\begin{figure*}
\includegraphics[width=0.93\textwidth]{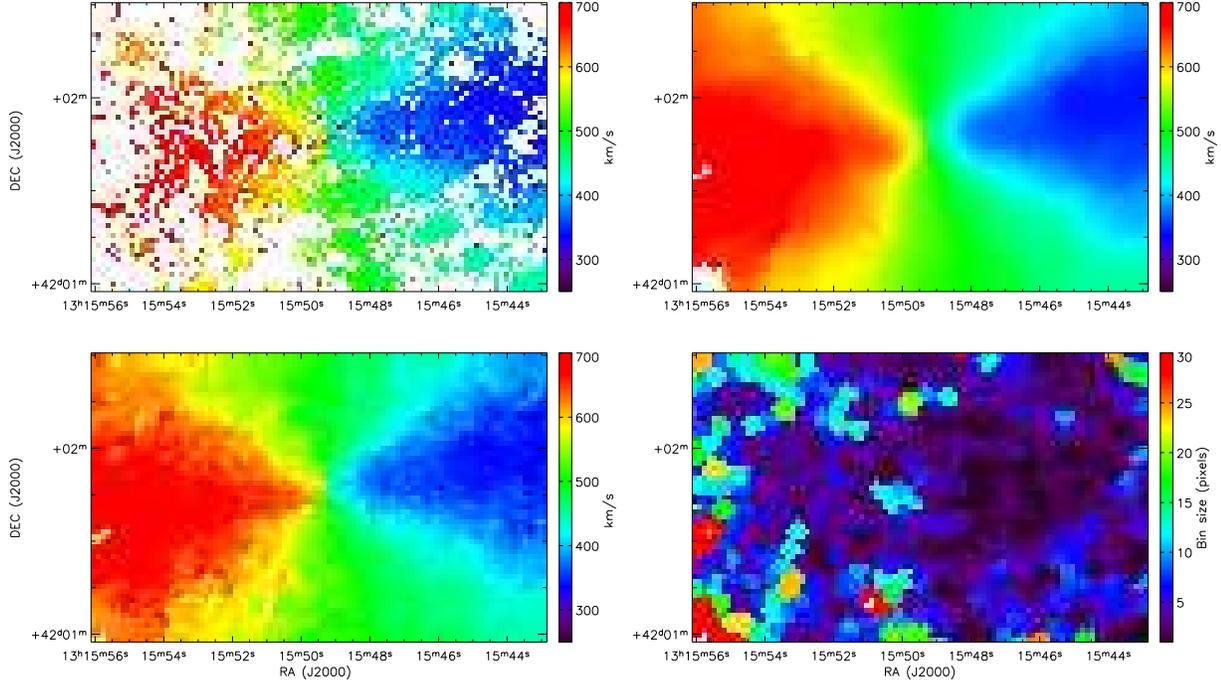}
\caption{Comparison of the 2D rotation velocity maps of \mbox{NGC 5055} at the centre of the galaxy with different smoothing algorithms. \mbox{\textbf{Top left} :} No smoothing. \mbox{\textbf{Top right} :} Gaussian 6x6 kernel. \mbox{\textbf{Bottom left} :} Adaptive smoothing, target SNR of 7. \mbox{\textbf{Bottom right} :} Bin sizes obtained with the adaptive smoothing.}
\label{figure_comp_voro_2d_centre}
\end{figure*}

\begin{figure*}
\includegraphics[width=0.93\textwidth]{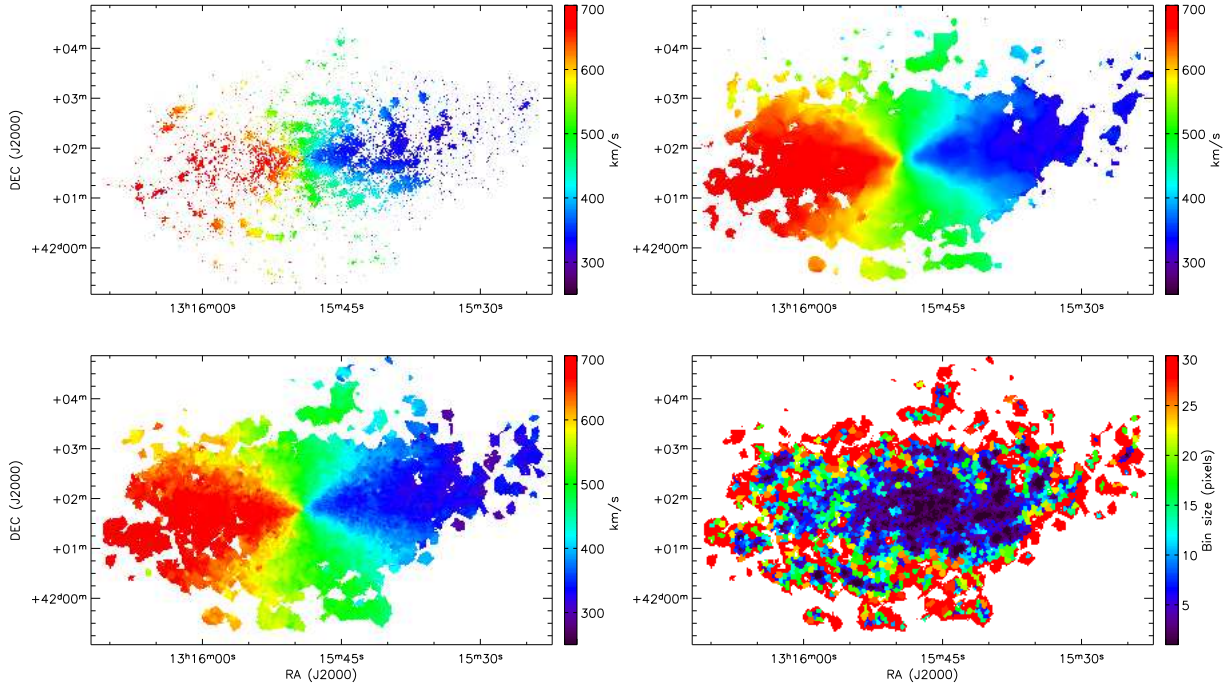}
\caption{Comparison of the full 2D rotation velocity maps of \mbox{NGC 5055} with different smoothing algorithms. \mbox{\textbf{Top left} :} No smoothing. \mbox{\textbf{Top right} :} Gaussian 6x6 kernel. \mbox{\textbf{Bottom left} :} Adaptive smoothing, target SNR of 7. \mbox{\textbf{Bottom right} :} Bin sizes obtained with the adaptive smoothing.}
\label{figure_comp_voro_2d}
\end{figure*}

\subsection{Noise sources and SNR}
Due to the absence of read-out noise, FP observations at \ha\, in photon counting mode are dominated by photon noise. One component of the SNR must thus be the noise of the sum of the photons (N) contained in the emission line that are above the continuum (the photons that create the emission line itself). Also, to account for the improperly removed sky lines, it is assumed that the continuum should be flat and its dispersion ($\sigma$) is used as another source of noise. This dispersion can also account for the read-out noise of the detector while processing CCD data. Obviously, in this case, the CCD signal must be gain-corrected so that the data is expressed in photons. The SNR is computed as follows
$$SNR = \frac{N}{\sqrt{N+\sigma^{2}}},$$
where
$$N = \sum_{i=0}^{i=n-1}\left(y_{i} - c\right)$$
is the number of photons in the emission line and $c$ is the continuum level.

\subsection{Binning and smoothing}
The binning algorithm used is an improvement of the one defined by \citeauthor{2003MNRAS.342..345C} (\citeyear{2003MNRAS.342..345C}), with the difference here that it allows the SNR of the bin to be recalculated every time a new spectrum is added. This method leads to a greater scatter in the resulting SNR of the bins, but, as previously stated, it allows the limitation of the spatial resolution to be taken into account. This means that the SNR of the bins will not necessarily grow as more spectra are added. It was found that targeted SNR values in the range of 5--10 are optimal (SNR values higher than 10 creating bins that are too large and into which the SNR is not increasing as more pixels are added). The original algorithm was also optimised to allow the processing of large data cubes (typically 512$\times$512 pixels and 48 channels) in a reasonable amount of time (e.g. less than 10 minutes on a typical personal computer).

After the bins are delimited, to avoid singularities at the bins boundaries, a Delaunay triangulation algorithm is used to smooth the edged of the bins, taking the flux weighted centroid of the bins as the nodal points of the triangles. This smoothed data cube is finally used to extract the radial velocity map, using the algorithm described in section \ref{section_finding_emission} to find the barycentre of the emission line.

Applying an accurate sky emission subtraction is very important before attempting to adaptively smooth a data cube. As the SNR of a spectrum is calculated by looking at only one emission line, an inadequately subtracted sky will cause the algorithm to trigger on the sky residual rather than on the galaxy's signal. The resulting smoothed cube will therefore have a constant SNR on the sky residual and the SNR of the galaxy's emission will rarely be optimum.


\section{Results}
\label{section_results}

Figures \ref{figure_comp_voro_2d_centre} and \ref{figure_comp_voro_2d} show the results of different levels of smoothing on the extracted velocity maps of the galaxy \mbox{NGC 5055}, using \ha. Figure \ref{figure_comp_voro_2d_centre} concentrates on the central parts of the galaxy, where the accuracy of the kinematical information is crucial (see section \ref{results_rc}). The high resolution (non smoothed) velocity map gives accurate kinematical information but has a very small spatial coverage. When a Gaussian smoothing is applied, these small kinematical details are lost, rendering the radial velocity map inaccurate. Note that the unsmoothed map has been thresholded using its associated monochromatic map to avoid having pixels of low monochromatic flux, whose velocities were incorrectly determined, polluting the map. A threshold of 5 photons has been used.

When considering the extended part of the galaxy (Figure \ref{figure_comp_voro_2d}), it is obvious that spatial smoothing must be applied in order to resolve the kinematical information at large galactocentric radii. In that case, a strong smoothing permits to have the best signal coverage but still causes velocity inaccuracies in the high SNR regions.

\begin{figure}
\includegraphics[width=0.49\textwidth]{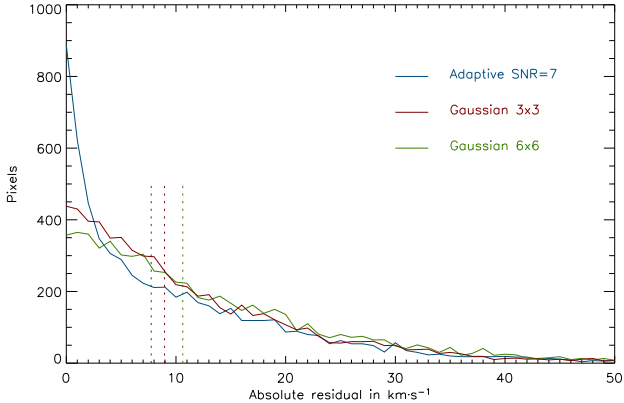}
\caption{Histograms of the absolute velocity residuals of the velocity maps (smoothed -- non-smoothed) of \mbox{NGC 5055} obtained with different smoothing algorithms. The vertical lines show the median residual for each algorithm. For the Gaussian 6x6 and 3x3 algorithms, the median residuals are 10.7 \kms\, and 8.9 \kms\, respectively. The median residual for the adaptive smoothing is 7.7 \kms. Note how the histogram of the residual of the adaptive smoothing peaks towards 0 \kms.}
\label{figure_comp_voro_2d_residu_histogram}
\end{figure}

In both cases, the adaptively smoothed map gives the best results: high resolution, high level of detail, great coverage at the centre of the galaxy and wide signal coverage in the low SNR regions. The median residuals of the adaptively smoothed map (Figure \ref{figure_comp_voro_2d_residu_histogram}) are lower than the ones obtained with other smoothing algorithms (7.7 \kms\, against 8.9 \kms\, and 10.7 \kms for the Gaussian 3x3 and 6x6 smoothing). 

Naturally, this method is efficient when the SNR varies greatly from one region to another. If the SNR is the same everywhere, the effect of the adaptive smoothing will be roughly the same as a fixed size kernel smoothing.

\subsection{Effect on rotation curves}
\label{results_rc}

\begin{figure}
\includegraphics[width=0.49\textwidth]{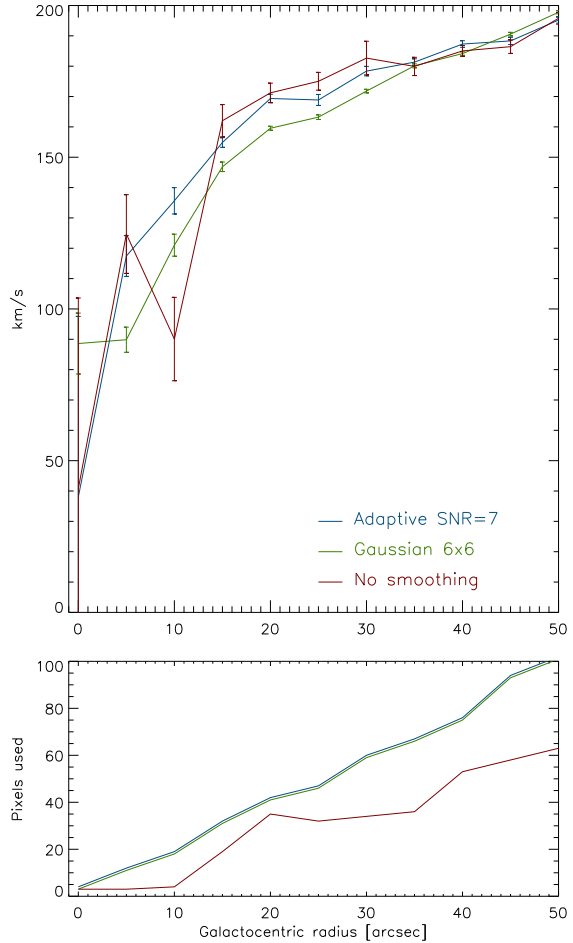}
\caption{Comparison of different smoothing algorithms applied to a galaxy velocity map and the result on the inner part of the extracted rotation curve. \mbox{\textbf{Top} :} The rotation curves. Note that the adaptive smoothing algorithm minimizes the beam smearing effect. \mbox{\textbf{Bottom} :} The number of points of the velocity map used in each crown of the tilted ring model to extract the rotation curve. Note that even at the centre of the galaxy, very few points are available in the non smoothed map.}
\label{figure_rc_centre}
\end{figure}

\begin{figure*}
\includegraphics[width=0.99\textwidth]{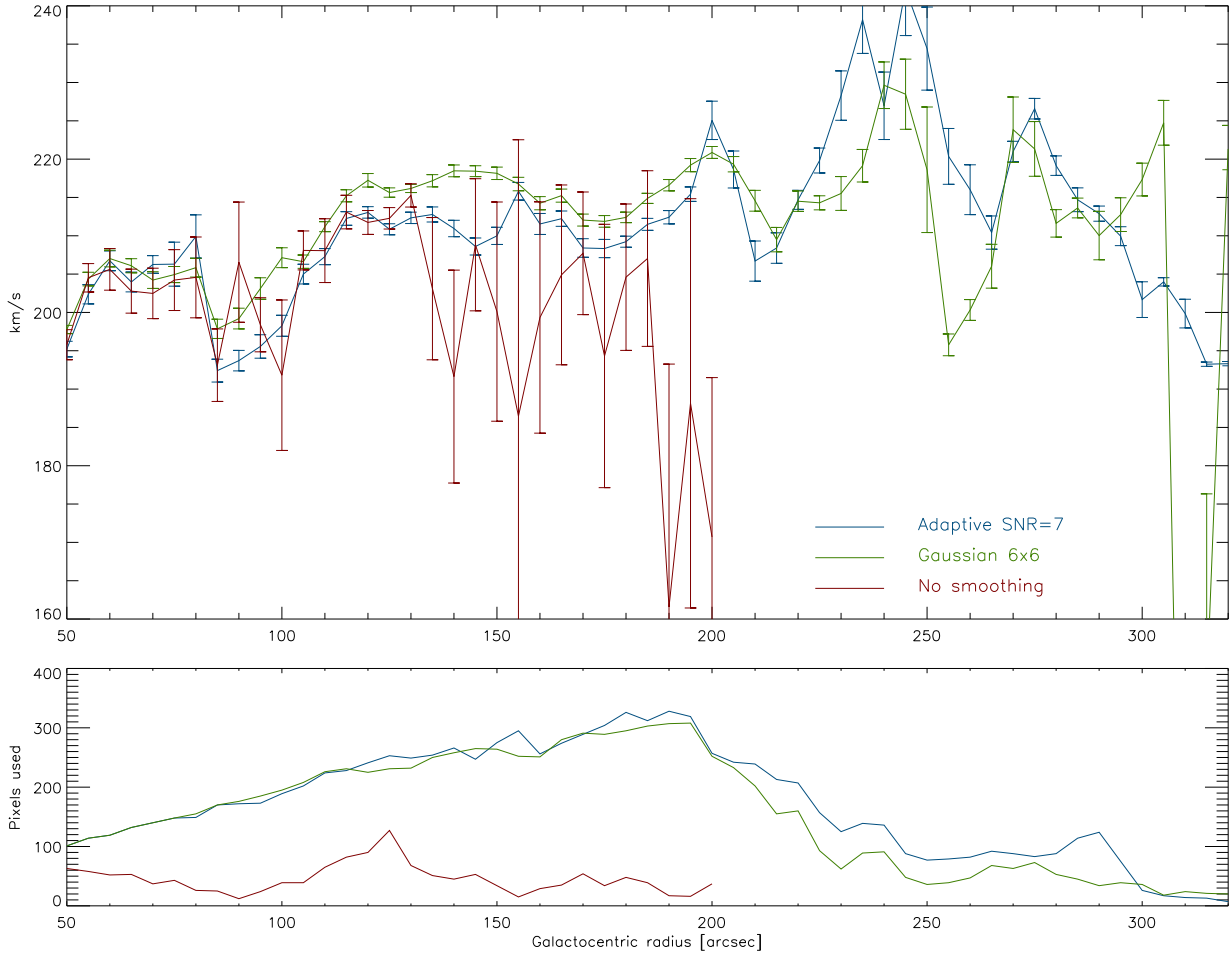}
\caption{Comparison of different smoothing algorithms applied to a galaxy velocity map and the result on the whole extent of the extracted rotation curve. Points beyond 200\Sec\, were removed for the non smoothed rotation curve as they were not accurate and were distracting. \textbf{Top} : The rotation curves. \textbf{Bottom} : The number of points of the velocity map used in each crown of the tilted ring model to extract the rotation curve.}
\label{figure_rc}
\end{figure*}

Rotation curves (RC) derived from the data cubes are shown in Figures \ref{figure_rc_centre} and \ref{figure_rc}. The RC were extracted using the \verb|rotcur| (\citeauthor{1987PhDT.......199B} \citeyear{1987PhDT.......199B}) routine of the \textsc{GIPSY} package, using fixed inclination (\textit{i}) and position angles (PA) at all radii. The error bars are the rotation velocity errors as determined by \verb|rotcur|. When there is little kinematical information into a ring or when there is a great scatter in the velocities composing the ring, the errors are large. The deep notch in the non smoothed rotation curve around 10\Sec\, is explained by the few emission points in the ring at that radius, as shown in the bottom panel of the figure, and is thus inaccurate. When a slight smoothing is applied, this feature disappears. However, when the smoothing is too strong as it is the case for a 6x6 Gaussian kernel, the rising part of the rotation curve is lowered. This produces an effect similar to the beam smearing observed in \hI\, data due to the limited spatial resolution. Adaptive smoothing should thus be preferred since, as shown by \citeauthor{1999AJ....118.2123B} (\citeyear{1999AJ....118.2123B}), the rising part of the rotation curve is crucial in the determination of the dark-to-luminous mass ratio and of the mass models parameters. Note that the rotation curve extracted from the adaptively smoothed radial velocity map is very close to the rotation curve extracted from the unsmoothed one within their error range ($\pm1 \sigma$) for nearly all points.

When considering the extended parts of the rotation curve, strong smoothing must be applied to resolve the kinematics, as the signal in these regions is very weak. Figure \ref{figure_rc} shows that very little kinematical information is available when no smoothing is applied. Thus, when considering fixed-size kernel smoothing, multiple rotation curves must be merged together to recover both resolution and signal coverage. Also, the Gaussian smoothing sometimes makes small detail of the RC disappear by convolving uncorrelated data points (like the structure around the galactocentric radius of 150\Sec which corresponds to a spiral arm of \mbox{NGC 5055}). The fall off seen for the adaptively smoothed curve at radii $>$ 250\Sec can be explained if the disk of \mbox{NGC 5055} is warped (\citeauthor{1981AJ.....86.1791B} \citeyear{1981AJ.....86.1791B}). This has not been modelled here.

The advantage of using an adaptive binning/smoothing algorithm is obvious for rotation curves extraction. It is the algorithm that produces the most accurate rotation curves, based on the largest spatial coverage possible. Moreover, it allows rendering high resolution kinematical information at the very centre of the galaxies where even an unsmoothed radial velocity map could not produce accurate results. Having a data cube that is smoothed to a constant SNR would also allow one to search for multiple emission line components with greater ease.

\section{Conclusion}
It has been shown that rigourous data reduction techniques applied to 3D FP data help getting as much information as possible out of data cubes with the best possible accuracy. Whilst good sky emission removal methods are necessary to avoid contaminating the weak galaxy's signal, a smoothing algorithm that adapts itself to the signal available helps preserving spatial resolution and optimises signal coverage. The effects of these methods are of high scientific importance: they help providing accurate and extended kinematical information, as demonstrated in section \ref{results_rc}.

It is clear when comparing the three smoothing schemes in Figures \ref{figure_comp_voro_2d_centre} and \ref{figure_comp_voro_2d} that the adaptive smoothing is doing as well as the heavy (6x6) Gaussian smoothing in the regions of weak signals while preserving the kinematical details in the regions of strong signals. Moreover, this technique avoids underestimating the velocities (beam smearing effect) in the inner parts of the rotation curves where the velocity gradient is strong, thus providing much more accurate kinematical data.

These data reduction techniques have already been applied for data presented in survey observation papers, such as the Big \ha\, kinematics sample of BARred spiral galaxies (\citeauthor{2005MNRAS.360.1201H} \citeyear{2005MNRAS.360.1201H}), the \ha\, Kinematics of the SINGS Nearby Galaxies Survey (\citeauthor{2006MNRAS.367..469D} \citeyear{2006MNRAS.367..469D}) and the Virgo high-resolution \ha\, kinematical survey (\citeauthor{2005astro.ph.11417C} \citeyear{2005astro.ph.11417C}). The data for these surveys were gathered at the Observatoire du mont M\'egantic 1.6-m telescope, the Canada--France--Hawaii 3.6-m telescope and the ESO/La Silla 3.6-m telescope. Not only do these methods provide data of high quality, they also reduce the time spent processing the data itself. In fact, a great deal of time was needed to manually clean up radial velocity maps where the sky emission had been improperly removed. With the growing number of galaxies observed in \ha\, surveys, these techniques helps the processing of extensive amounts of data in a short period of time.

The data reduction techniques presented in this paper focus mostly on FP data obtained with an Image Photon Counting System at \ha. However, these methods can also be applied to other data sets, in particular to any 3D integral field data obtained with spectrometers based on grisms systems (optical fibres, microlenses, slicers). All the techniques described could also be applied to CCD FP data cubes. The sky emission subtraction method described in section \ref{section_sky_emission} should work for any data cube in which sky-dominated regions are apparent. In the case where the galaxy occupies the whole field of view, an observation of the sky must be interlaced with the observation of the galaxy and the sky cube may then be subtracted from the galaxy cube.

\section*{Appendix I: Availability}
The data reduction routines described throughout this paper are available at \url{http://www.astro.umontreal.ca/fantomm/reduction} in IDL language. The intensity weighted mean algorithm has been written in C language for performance reasons. The C module is loaded in IDL through the Dynamically Lodable Modules (DLM) mechanism. A Makefile is provided so that the compilation of the module under different flavours of Unix operating systems should be made easy (including Linux and Mac \mbox{OS X}). The routines are also working in IDL under Windows. The routines can use as input files raw observational data (usually ADA files, containing only photon positions for the observation and AD3 files for the calibrations. See \url{http://www-obs.cnrs-mrs.fr/adhoc/adhoc.html} for AD* files format.) or they can be feed with interferograms cubes that have not yet been phase-corrected, phase-corrected cubes, phase-corrected and sky-subtracted cubes, and so on. This makes it possible to process data using another data reduction package and use these routines for some phases of the data reduction only. Other file types (such as the more standard FITS format) can be supported with the supplied FITS to AD* and AD* to FITS converters. The software also provides mosaicing and merging facilities to allow for large galaxies and galaxies observed at different times to be processed. This package does not (yet) include visualisation routines or any kind of interactive user interface. Visualisation applications of the ADHOCw package can be used (\url{http://www-obs.cnrs-mrs.fr/adhoc/adhoc.html}) to view AD* files. They are designed to run under the Windows operating system, but they have been reported to work fine under Linux using the Windows Emulator (WINE) package (\url{http://www.winehq.com/}).

\section*{Appendix II: Observational parameters}
The observational data used in this paper were gathered with \FM\, (\citeauthor{2003SPIE.4841.1472H} \citeyear{2003SPIE.4841.1472H}, \citeauthor{2002PASP..114.1043G} \citeyear{2002PASP..114.1043G}), an integral field photon counting spectrometer using a Fabry-Perot etalon. The observational parameters for galaxy used as case studies in this paper are presented in Table \ref{observational_data}.

\begin{table}
\caption{Observational parameters} 
\label{observational_data}
\begin{tabular}{lcc}
\hline
\hline
Parameter & \mbox{NGC 5055} & \mbox{NGC 5194}\\
\hline
Type                            & SA(rs)bc       & SA(s)bc pec \\
$\mathrm{V_{sys}}$ (\kms)  & 504  & 463 \\
Observation date                & 14/03/2004     & 18/05/2003  \\
Field $\mathrm{RA_{central}}$ (J2000) & 13h15m58.3 s   & 13h30m07.3s \\
Field $\mathrm{DEC_{central}}$(J2000) & +42d00m11.3s   & +47d12m11.0s \\
Field of view (\Min)            & 13.8 x 13.8 & 13.8 x 13.8 \\
Pixel size (\Sec)               & 1.61 x 1.61 & 1.61 x 1.61 \\
Focal ratio  (F/D)              & 2.0      & 2.0\\
Filter $\lambda_{central}$ (\AA) & 6584    & 6581 \\
Filter FWHM (\AA)               & 15.1 & 19.8 \\
Filter transmission  (\%)     & 74   &  60 \\
FP interference order (p)           & 765  & 899 \\
FP FSR (\kms)        & 391.88 & 333.47 \\
Number of channels                  & 48   & 48 \\
Channel step (\AA)                  & 0.15 & 0.18 \\
Calibration Finesse                  & 16.8 & 20.7 \\
Resolution ($\frac{\delta\lambda}{\lambda}$) & 12852 & 18609\\
Total exposure (min)           & 128  &  246 \\
Exposure per channel (min)     & 2.66 & 5.14 \\
\hline

\end{tabular}
\end{table}



\section*{Acknowledgements}

This study has been funded by the Fond Qu\'eb\'ecois de Recherche sur la Nature et les Technologies, who partly supported Olivier Hernandez, Laurent Chemin and Olivier Daigle. Claude Carignan's work was funded by the Conseil National de Recherche en Sciences Naturelles et G\'enie du Canada.

The \FM\, project has been carried out by the Laboratoire d'Astrophysique 
Exp\'erimentale (\textbf{\texttt{LAE}}) of the Universit\'e de Montr\'eal using a 
grant from the Canadian Foundation for Innovation and the Minist\`ere de l'\'Education 
du Qu\'ebec.

We thank the anonymous referee for his valuable comments.

\bsp

\label{lastpage}

\end{document}